# Macroscopic Rabi-Like Oscillations of Ultracold Atoms in an Asymmetrical Two-Dimensional Magnetic Lattice

**A. Abdelrahman[1], M. Vasiliev[1] and K. Alameh[1,2]**

[1] *Electron Science Research Institute, Edith Cowan University*
*270 Joondalup Drive, Perth WA 6027 Australia*

[2] *WCU Department of Nanobio Material and Electronics, Gwangju Institute of Science and Technology, Gwangju 500-712, South Korea*

*a.abdelrahman@ecu.edu.au*

**Abstract:** We investigate the existence of the *macroscopic quantum phase* in trapped ultracold quantum degenerate gases, such as Bose-Einstein condensate, in an asymmetrical two-dimensional magnetic lattice. We show the key to adiabatically control the tunneling in the new two-dimensional magnetic lattice by means of external magnetic bias fields. The macroscopic quantum phase signature is identified as a Rabi-like oscillation when solving the system of coupled time-dependent differential equations, described here by the Boson Josephson Junctions (BJJs). In solving the system of the BJJs we used an order parameter that includes both time-dependent variational parameters which are the fractional population at each lattice site and the phase difference. The BJJs solution presents a clear evidence for the macroscopic quantum coherence.

---

**References and links**

1. K. Mahmud, H. Perry, and W. Reinhardt, "*Quantum phase-space picture of Bose-Einstein condensates in a double well*", Phys. Rev. A **71**, 023615 (2005).
2. M. Andrews, C. Townsend, H. Miesner, D. Durfee, D. Kurn, and W. Ketterle, "*Observation of Interference Between Two Bose Condensates*", Science **275**, 637 (1997).
3. D. Hall, M. Matthews, C. Wieman, and E. Cornell, "*Measurements of Relative Phase in Two-Component Bose-Einstein Condensates*", Phys. Rev. Lett. **81**, 1543 (1998).
4. S. Papp, J. Pino, R. Wild, S. Ronen, C. Wieman, D. Jin, and E. A. Cornell, "*Bragg Spectroscopy of a Strongly Interacting $^{85}$**Rb** Bose-Einstein Condensate*", Phys. Rev. Lett. **101**, 135301 (2008).
5. N. Claussen, E. Donley, S. Thompson, and C. Wieman, "*Microscopic Dynamics in a Strongly Interacting Bose-Einstein Condensate*", Phys. Rev. Lett. **89**, 010401 (2002).
6. J. Estéve, C. Gross, A. Weller, S. Giovanazzi, and M. Oberthaler, "*Squeezing and entanglement in a Bose-Einstein condensate*", Nature **455**, 1216 (2008).
7. Q. Chena, J. Stajicb, S. Tanb, and K. Levinb, "*BCS-BEC crossover: From high temperature superconductors to ultracold superfluids*", Physics Reports **412**, 1 (2005).
8. A. Schirotzek, Y. Shin, C. Schunck, and W. Ketterle, "*Determination of the Superfluid Gap in Atomic Fermi Gases by Quasiparticle Spectroscopy*", Phys. Rev. Lett. **101**, 140403 (2008).
9. B. Josephson, "*Possible New Effects in Superconducting Tunnelling*", Phys. Rev. Lett. **1**, 251 (1962).
10. S. Levy, E. Lahoud, I. Shomroni, and J. Steinhauer, "*The a.c. and d.c. Josephson effects in a Bose-Einstein condensate*", Nature **449**, 579 (2007).
11. G. Milburn, J. Corney, E. Wright, and D. Walls, "*Quantum dynamics of an atomic Bose-Einstein condensate in a double-well potential*", Phys. Rev. A **55**, 4318 (1997).

12. R. Spekkens and J. Sipe, "*Spatial fragmentation of a Bose-Einstein condensate in a double-well potential*", Phys. Rev. A **59**, 3868 (1999).
13. I. Bloch, "*Ultracold quantum gases in optical lattices*", Nature. Phys. **1**, 23 (2005).
14. B. Paredes, A. Widera, V. Murg, O. Mandel, S. Fölling, I. Cirac, G. V. Shlyapnikov, T. W. Hänsch, and I. Bloch, "*Tonks-Girardeau gas of ultracold atoms in an optical lattice*", Nature (London) **429**, 277 (2004).
15. Z. Hadzibabic, P. Krüger, M. Cheneau, B. Battelier, and J. Dalibard, "*Berezinskii-Kosterlitz-Thouless crossover in a trapped atomic gas*", Nature **441**, 118 (2006).
16. O. Mandel, M. Greiner, A. Widera, T. Rom, W. Hänsch, and I. Bloch, "*Coherent Transport of Neutral Atoms in Spin-Dependent Optical Lattice Potentials*", Phys. Rev. Lett. **91**, 010407 (2003).
17. M. Greiner, O. Mandel, T. W. Hänsch, and I. Bloch, "*Collapse and revival of the matter wave field of a Bose-Einstein condensate*", Nature **51**, 419 (2002).
18. R. Jördens, N. Strohmaier, K. Günter, H. Moritz, and T. Esslinger, "*A Mott insulator of fermionic atoms in an optical lattice*", Nature **455**, 204 (2008).
19. J. Reichel, "*Microchip traps and Bose-Einstein condensation*", Appl. Phys. B **74**, 469 (2002).
20. H. Ott, J. Fortágh, G. Schlotterbeck, A. Grossmann, and C. Zimmermann, "*Bose-Einstein Condensation in a Surface Microtrap*", Phys. Rev. Lett. **87**, 230401 (2001).
21. W. Hänsel, P. Hommelhoff, T. W. Hänsch, and J. Reichel, "*Bose-Einstein condensation on a microelectronic chip*", Nature **413**, 498 (2001).
22. J. Fortágh and C. Zimmermann, "*Magnetic microtraps for ultracold atoms*", Rev. Mod. Phys. **79**, 235 (2007).
23. P. Treutlein, P. Hommelhoff, T. Steinmetz, T. Hänsch, and J. Reichel, "*Coherence in Microchip Traps*", Phys. Rev. Lett. **92**, 203005 (2004).
24. A. Abdelrahman, P. Hannaford, and K. Alameh, "*Adiabatically induced coherent Josephson oscillations of ultracold atoms in an asymmetric two-dimensional magnetic lattice*", Optics Express **17**, 24358 (2009).
25. A. Abdelrahman, P. Hannaford, M. Vasiliev, and K. Alameh, "*Asymmetrical Two-dimensional Magnetic Lattices for Ultracold Atoms*", Phys. Rev. A (in press) (2009).
26. M. Singh, R. McLean, A. Sidorov, and P. Hannaford, "*Dynamics of reflection of ultracold atoms from a periodic one-dimensional magnetic lattice potential*", Phys. Rev. A **79**, 053407 (2009).
27. S. Whitlock, R. Gerritsma, T. Fernholz, and R. Spreeuw, "*Two-dimensional array of microtraps with atomic shift register on a chip*", New J. Phys. **11**, 023021 (2009).
28. P. Anderson, "*Considerations on the Flow of Superfluid Helium*", Rev. Mod. Phys. **38**, 298 (1969).
29. S. Raghavan, A. Smerzi, S. Fantoni, , and S. R. Shenoy, "*Coherent oscillations between two weakly coupled Bose-Einstein condensates: Josephson effects, $\pi$-oscillations, and macroscopic quantum self-trapping*", Phys. Rev. A **59**, 620 (1999).
30. A. Smerzi, S. Fantoni, S. Giovanazzi, and S. Shenoy, "*Quantum Coherent Atomic Tunneling between Two Trapped Bose-Einstein Condensates*", Phys. Rev. Lett. **79**, 4950 (1997).
31. I. Zapata, F. Sols, and A. Leggett, "*Josephson effect between trapped Bose-Einstein condensates*", Phys. Rev. A **57**, R28 (1998).
32. S. Giovanazzi, A. Smerzi, and S. Fantoni, "*Josephson Effects in Dilute Bose-Einstein Condensates*", Phys. Rev. Lett. **84**, 4521 (2000).
33. A. Smerzi, S. Fantoni, and S. Giovanazzi, "*Macroscopic Quantum Coherence Phenomena in Bose-Einstein Condensates*", Sissa Digital Library (Italy), PhD thesis (1998).
34. S. Inouye, M. Andrews, J. Stenger, H. Miesner, D. Stamper-Kurn, and W. Ketterle, "*Observation of Feshbach resonances in a Bose-Einstein condensate*", Nature **392**, 151 (1998).
35. F. Cataliotti, S. Burger, C. Fort, P. Maddaloni, F. Minardi, A. Trombettoni, A. Smerzi, and M. Inguscio, "*Josephson Junction Arrays with Bose-Einstein Condensates* ", Science **293**, 843 (2001).

## 1. Introduction

The *macroscopic quantum phase* signature exhibited by quantum degenerate gases at low temperature, such as Bose-Einstein condensates (BECs), remarkably has presented a detectable inherited coherence [1]. Such existing nature of *macroscopically* pronounced quantum coherence has triggered an intensive search to identify critical phase transitions. As a consequence, it is now feasible to detect the interference patterns, i.e. accessible interference fringes [4], of overlapped quantum degenerate gases [2][3]. Such coherent coupling allowed close investigations of *strongly* interacting condensates using traditional spectroscopic techniques, such as Bragg spectroscopy [5]. Moreover, the macroscopic quantum phase is also used to classically map a rich interacting nature of *weakly* coupled reservoirs of ultracold quantum degenerate gases, e.g. BECs in double wells [1][6]. An environment with such signature provides a unique analogy to condensed matter systems with a capability to adiabatically control critical phase transitions, such as superconductivity [7] and superfluidity [8]. This made it possible to simulate a diverse range of interesting phenomena, for example the Josephson phenomena [9][10], the deBroglie wave interference [2] and Feshbach resonance [34] . The essential underlying physics of the macroscopic finger print in such systems can be understood by studying a simplified model of a double potential well BEC with an adiabatically controlled barrier height using the well known quantum two-mode approximation [11][12].

Due to the fragility of macroscopically inferred quantum patterns, specialized experimental setups are required to bring specific interactions to adiabatic stage. Remarkable results have been achieved using optical lattices [13]; they are able to coherently manipulate trapped low-dimensional quantum gases [14][15] and transfer them, via spin-dependent mechanisms, between the lattice sites [16]. Their classical coherent mapping has enabled simulating several critical phase transitions of condensed matter systems such as Mott-insulator-to-superfluid transition by using both trapped BECs [17] and trapped ultracold fermionic gases [18].

On the other hand, integrating ultracold atoms with magnetic microstructure [19] and the creation of Bose-Einstein condensate on-atom-chip [20][21][22] have triggered an alternative approach to adiabatically inducing an identity for macroscopic quantum phases. Record breaking 1 second of coherence lifetime has been achieved using magnetic microchip [23]. In this article we present a possible realization of the macroscopic quantum phase signature, using our recent development [24][25] in the field of magnetic lattices [26][27].

## 2. The Two-Dimensional Magnetic Lattice Structure and The Adiabatic Tunneling Mechanism

Asymmetrical two-dimensional magnetic lattices [24][25] are realized by milling a $n \times n$ array of square holes of size of $\alpha_h \times \alpha_h$ and separated by distances of width $\alpha_s$ where we assume in our analysis an infinite lattice with $\alpha_h = \alpha_s \equiv \alpha$. The structure is formed in a magneto-optic thin film of thickness $\tau$ as shown in Figure 1(a). The presence of the distributed $n \times n$ holes in a magnetized thin film results in spatially distributed magnetic field minima having the same periodic pattern and located at effective distances, $d_{min}$, from the top of the holes as shown in Figure 2(a). The spatial magnetic field components $B_x$, $B_y$ and $B_z$ are written in terms of a field decaying away from the surface of the trap in the $z$-direction, produced by magnetic induction, $B_o = \mu_o M_z / \pi$, and distributed periodically in the $x-y$ plane at the surface of the magnetized thin film. $B_o$ is used to derive a surface reference magnetic field defined as $B_{ref} = B_o(1 - e^{-\beta \tau})$, with $\beta = \pi / \alpha$. Assuming a plane of symmetry at $z = 0$ we write an analytical expression for the spatially distributed trapping magnetic field $B(x,y,z)$ which also includes external magnetic bias field components, $B_{x-bias}$, $B_{y-bias}$ and $B_{z-bias}$, as follow

$$B(x,y,z) = \left\{ B_{x-bias}^2 + B_{y-bias}^2 + B_{z-bias}^2 + 2B_{ref}^2 \left[1 + cos(\beta x)cos(\beta y)\right] e^{-2\beta[z-\tau]} + \right.$$
$$\left. + 2B_{ref} e^{-\beta[z-\tau]} \left( sin(\beta x) B_{x-bias} + sin(\beta y) B_{y-bias} + \left[cos(\beta x) + cos(\beta y)\right] B_{z-bias} \right) \right\}^{1/2}$$
(1)

The details of relevant derivations were reported in [25]. The heights of the tunneling barriers $\Delta B(\mathbf{x})$ of each individual potential well, which represents a lattice site, are determined by their minima and maxima as

$$\Delta B(\mathbf{x}) = |B_{max}(\mathbf{x})| - |B_{min}(\mathbf{x})|, \qquad \mathbf{x} \equiv (x,y,z) \tag{2}$$

It is of particular importance to carefully determine the curvature of the trapping magnetic field at each individual site and that is because lattice sites with steeper gradients may individually develop Majorana spin-flips destructive process. The curvatures of the magnetic field along the confining directions determine the trapping frequencies, $\omega_k$ with $k \in \{x,y,z\}$, which depend on the Zeeman sub-levels. For the case of a harmonic potential the frequencies $\omega_k$ are given by

$$\omega_k = \frac{\beta}{2\pi}\sqrt{\mu_B g_F m_F \frac{\partial^2 B}{\partial k^2}} \qquad k \equiv x,y,z \qquad (3)$$

where $g_F$ is the Landé g-factor, $\mu_B$ is the Bohr magneton, and $m_F$ is the magnetic quantum number of the hyperfine state. We find that at the centers of the traps $\frac{\partial^2 B}{\partial x^2} = \frac{\partial^2 B}{\partial y^2}$ holds due to the symmetry. Also, using the value of the non-zero local minimum $B_{min}$ the depth of the magnetic lattice sites $\Lambda_{depth}$ can be determined as follow

$$\Lambda_{depth}(\mathbf{x}) = \frac{\mu_B g_F m_F}{k_B}\Delta B(\mathbf{x}) \qquad (4)$$

where $k_B$ is the Boltzmann constant. Figures 1(b) shows a 3D plot of the distributed magnetic field at the effective distance $d_{min}$ which is found to be $d_{min} > \alpha/2\pi$ [25]. Simulated maps of the magnetic field strength distribution across the $x-y$ plane located at $d_{min}$ above the magnetized film surface are shown in Figures 1(c,d) with and without the applications of the external bias fields $B_{x-bias}$ and $B_{y-bias}$, respectively where it is clear that the bias fields have no effects on the sites geometry at the center of the magnetic lattice.

Figure 2(a) shows the asymmetrical behavior realized in this new type of magnetic lattices which can be used to introduce adiabatically controlled gravitational offsets (space gaps) along the gravitational field $z$-axis [28].

The key to the adiabatically controlled tunneling is demonstrated in Figures 2(b,c) where the application of an external $z$-bias field, $B_{z-bias}$, significantly influences the value of $d_{min}$ and $B_{min}$ and eventually the magnetic states of the trapped atoms at each site thus leading to desirably controlled oscillations of the ground and the excited states (or superposition states) between the lattice sites [24]. It is important to keep $d_{min}$ sufficiently large, so as to keep the trapped cold atoms away from the surface Casimir-Polder interacting limits; this can initially be well maintained via $B_{z-bias}$.

In the following section, for simplicity, we only consider a one-directional tunneling starting from the center of the trap as shown in Figure 2(d.1). The asymmetrical effect is used to determine the tunneling directions of the trapped ultracold atoms across the $x-y$ and $x,y-z$ planes as shown in Figure 2. Measurement of the confining magnetic field is reported in [24][25].

## 3. Bosonic Josephson Junctions in the 2D Magnetic Lattice

The time-dependent variational wavefunction, $\varphi(\mathbf{x},t) = \sum_i^n c_i(t)\chi_i(\mathbf{x})$ for $n$ lattice sites, is used to roughly describe the *macroscopic* dynamical evolution of the weakly interacting trapped condensates in the magnetic lattice obeying the time-dependent Gross-Pitaevskii equation (GPE), i.e. the non-linear Schrödinger equation

$$i\hbar\partial_t \varphi(\mathbf{x},t) = \left[-\frac{\hbar^2}{2M}\nabla^2 + U(\mathbf{x}) + g_o|\varphi(\mathbf{x},t)|^2\right]\varphi(\mathbf{x},t) \qquad (5)$$

where $g_o = \frac{4\pi\hbar^2 a_s}{M}$ is the inter-atomic scattering pseudo potential with $a_s$ the s-wave scattering length and $M$ is the atomic mass. $U(\mathbf{x})$ is the external harmonic trapping potential written as

$$U(\mathbf{x}) = \frac{1}{2}M\sum_{k\in\{x,y,z\}}\omega_k^2 k^2 + \delta z \qquad (6)$$

in which $\delta z$ determines the amount of tilt introduced in $U(\mathbf{x})$ and the trapping frequencies $\omega_k$ are calculated using Equation (4) with the external magnetic field $B$ as described in Equation (1).

For $n$ adjacent lattice sites, described as Boson Josephson Junctions (BJJs) [29][30], the system of time-dependent Schrödinger equations, i.e. the BJJs equations, is described by the following Hamiltonian matrix

$$\widehat{\mathscr{H}} = \begin{pmatrix} A_{1,1} & -\Omega^J_{1,2} & 0 & 0 & \ldots & 0 \\ -\Omega^J_{2,1} & A_{2,2} & -\Omega^J_{1,2} & 0 & \ldots & 0 \\ 0 & -\Omega^J_{3,2} & A_{3,3} & -\Omega^J_{2,3} & \ldots & 0 \\ \vdots & \vdots & \ddots & \ddots & \ddots & \vdots \\ 0 & \ldots & 0 & -\Omega^J_{n,n-1} & A_{n,n} & -\Omega^J_{n-1,n} \end{pmatrix} \quad (7)$$

where $A_{i,i} \equiv E^o_i + \Gamma_i N_i$. $E^o_i$ with $i \in \{1,n\}$ are the zero-point energies in each individual lattice site in which $E^o_i = \int [\frac{\hbar^2}{2M}|\nabla \varphi_i|^2 + |\varphi_i|^2 U(\mathbf{x})]d\mathbf{x}$. The atomic self-interaction energy is determined by $\Gamma_i = g_o \int |\varphi_i|^4 d\mathbf{x}$.

We consider a scenario where only the superposition of the wavefunctions is within each two adjacent lattice sites, in the one-dimension, and hence we write the coupling Josephson energy as $\Omega^J_{i,i+1} \simeq -\int [\frac{\hbar^2}{2M}(\nabla \varphi_i \nabla \varphi_{i+1}) + \varphi_i U(\mathbf{x})\varphi_{i+1}]d\mathbf{x}$ which is in analogous to Josephson coupling energy in a Superconducting Josephson Junction [10][31][32]. Numerical evaluation of $\Omega^J$ is shown in Figure (4) for both cases the zero biased, $B_{x-bias} = B_{y-bias} = B_{z-bias} = 0$ G, magnetic lattice with $\delta = 0$ and for the externally biased lattice $B_{x-bias} = B_{y-bias} = -B_{z-bias} = 10$ G and $\delta \neq 0$. Since our interest is to identify the dynamical Josephson oscillations of the condensates in the weakly coupled sites of the magnetic lattice, we thus limit our attention to the localized time-dependent variational ansatz $\varphi(\mathbf{x},t) = c_i(t)\chi_i(\mathbf{x}) + c_{i+1}(t)\chi_{i+1}(\mathbf{x})$ considering the total number of atoms between the adjacent sites $\mathcal{N} = N_i + N_{i+1} = |c_i|^2 + |c_{i+1}|^2$ to be constant. General occupation and localized phases obey the two-mode dynamical BJJs equations

$$i\hbar \partial_t c_i(t) = [E^o_i + \Gamma_i N_i]c_i(t) - \Omega^J_{i,i+1}c_{i+1}(t) \quad (8)$$
$$i\hbar \partial_t c_{i+1}(t) = [E^o_{i+1} + \Gamma_{i+1}N_{i+1}]c_{i+1}(t) - \Omega^J_{i+1,i}c_i(t) \quad (9)$$

The Dirac action principle leads to a classical expression of the system Hamiltonian, as explained in the following section, for two variational parameters described as the fractional population at each lattice site $\widetilde{N}(t) = \frac{N_i(t) - N_{i+1}(t)}{\mathcal{N}}$ and the phase difference $\widetilde{\theta}(t) = \theta_i(t) - \theta_{i+1}(t)$ in which both variables are canonical conjugates [33].

## 4. The non-interacting limit and the Rabi-like Oscillations

The time-dependent variational two-mode approximation can be used to describe the interaction dynamic in the $n \times n$ magnetic lattice where the set of the trail wavefunctions $\chi_i(t)$ in the order parameter $\varphi(\mathbf{x},t) = \sum_i^n c_i(t)\chi_i(\mathbf{x})$ span a subspace that constrains the system. Clearly, using the phase space conserved coordinates $c_i(t) = \sqrt{N_i(t)}e^{i\theta_i(t)}$, in which $\sum_i^n |c_i(t)|^2 = 1$, one realizes that $\widetilde{N}(t)$ and $\widetilde{\theta}(t)$ are the principal variables of the following non-interacting limit Hamiltonian

$$\widehat{\mathscr{H}}(\widetilde{N},\widetilde{\theta}) = E^o + \Omega^J \sqrt{1 - \widetilde{N}^2}\cos\widetilde{\theta} \quad (10)$$

where we considered the inter-atomic interactions to be negligible compared to the coupling Josephson energy, i.e. $\Gamma_i \ll \Omega^J_{i,i+1}$. Energy conservation per particle implies that the dynamics of the system is governed by a system of coupled equations which interestingly yields, for an exact solution, a Rabi-like oscillation. The system of the coupled time-dependent differential equations in the non-interacting limit can be written as

$$\dot{\widetilde{N}} \equiv \partial_{\widetilde{\theta}} \widehat{\mathscr{H}}(\widetilde{N}, \widetilde{\theta}) = \frac{2\Omega^J}{\hbar}\sqrt{1-\widetilde{N}^2}\sin(\widetilde{\theta}) \quad (11)$$

$$\dot{\widetilde{\theta}} \equiv \partial_{\widetilde{N}} \widehat{\mathscr{H}}(\widetilde{N}, \widetilde{\theta}) = \frac{-2\Omega^J}{\hbar}\frac{\widetilde{N}}{\sqrt{1-\widetilde{N}^2}}\cos(\widetilde{\theta}) \quad (12)$$

The system is solved in terms of $\widetilde{N}(t)$ and $\widetilde{\theta}(t)$, with a rescaled time $t \to \frac{2t}{\hbar}\Omega^J$, indicating the possibility of having a clear signature of the Rabi-like oscillations in this type of magnetic lattices [34]. Figures 3(a,b) show the oscillated modes $\widetilde{N}_i(t)$ and $\widetilde{\theta}_i(t)$ per site for different initial values. The oscillating fractional occupations $\widetilde{N}_{i,i+1}(t)$ and the phase amplitudes $\widetilde{\theta}_{i,i+1}(t)$ are shown in Figures 5(a,b), respectively, with initial conditions $\widetilde{N}_i(0) = 0.99$, $\widetilde{N}_{i+1}(0) = 0.5$ and $\widetilde{\theta}_i(0) = \widetilde{\theta}_{i+1}(0) = \pi$. These oscillations, with a frequency $\omega_R \approx |\Omega^J_{i,i+1}|$, are similar to the dynamical oscillations of single atoms however an adiabatic Josephson effect may also be realized in this type of magnetic lattice arising from the superfluidity nature of the trapped condensates [24][35].

## 5. Conclusion

We have shown a possibility of realizing a macroscopic quantum phase finger print in an asymmetrical two-dimensional magnetic lattice which is created using our recently-developed simple method. We have shown the key to adiabatically control atomic tunneling in the two-dimensional magnetic lattice by means of applying external magnetic bias fields. A Rabi-like oscillation has been identified by solving a system of coupled time-dependent differential equations describing the Boson Josephson Junctions (BJJs) in the magnetic lattice. An order parameter which includes the two time-dependent variational parameters, described here as the fractional population at each lattice site and their phase difference, has been used in solving the BJJs equations. The oscillations in the time-dependent variational parameters provide a clear evidence that the macroscopic quantum phase may exist in this type of magnetic lattice.

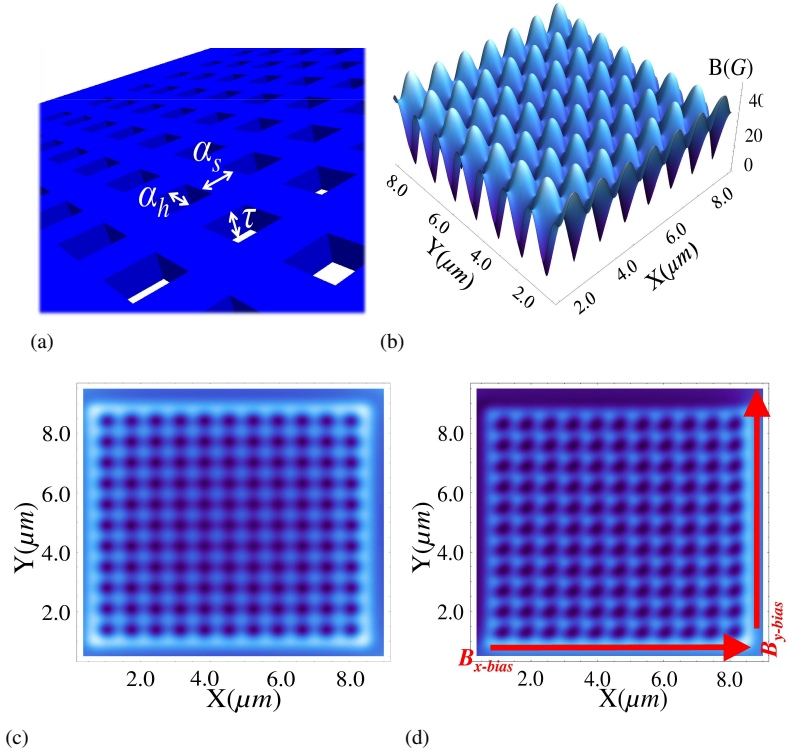

Fig. 1. (a) Lattice parameters are specified by the hole size $\alpha_h \times \alpha_h$, the separation between the holes $\alpha_s$ and the magnetic film thickness $\tau$. (b) 3D plot of the magnetic field of the distributed lattice sites across the $x-y$ plane at the effective $z$-distance, $d_{min}$. (c) Magnetic field density plots across the $x-y$ plane at $d_{min}$ with no external magnetic bias fields and (d) with the application of external bias fields $B_{x-bias} = B_{y-bias} = 10\ G$. External bias fields have slightly changed the magnetic lattice geometry. Simulation input parameters: $\alpha_s = \alpha_h = 1\ \mu m$, $M_z = 3\ kG$ and $\tau = 2\ \mu m$.

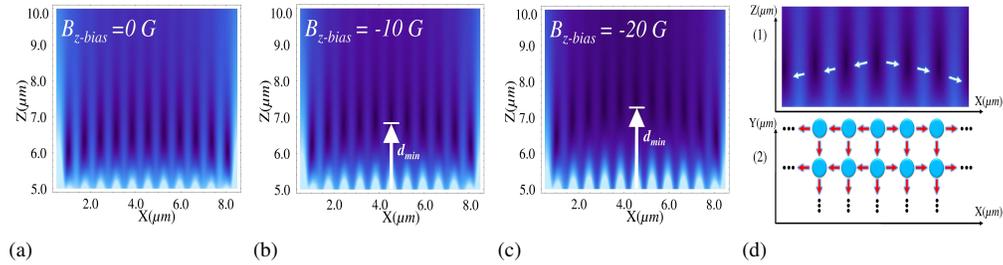

Fig. 2. (Color online) (a) Density plot representation of the simulated magnetic field of a finite magnetic lattice in the $z-x$ plane across the center of the traps where the asymmetrical effect is clearly present. (b-c) Effects of applying a $B_{z-bias}$ field, along the negative direction of the $z$-axis on the asymmetrical effect and the $d_{min}(\mu m)$. (b.1) Arrows indicate the tunneling directions of the ultracold atoms starting from the center of the trap. Diagonal tunneling may not be feasible in this type of magnetic lattice as schematically represented in (b.2). Simulation input parameters: $n = 11$ sites, $\alpha_s = \alpha_h = 3.5\ \mu m$, $M_z = 2.8\ kG$ and $\tau = 2\ \mu m$.

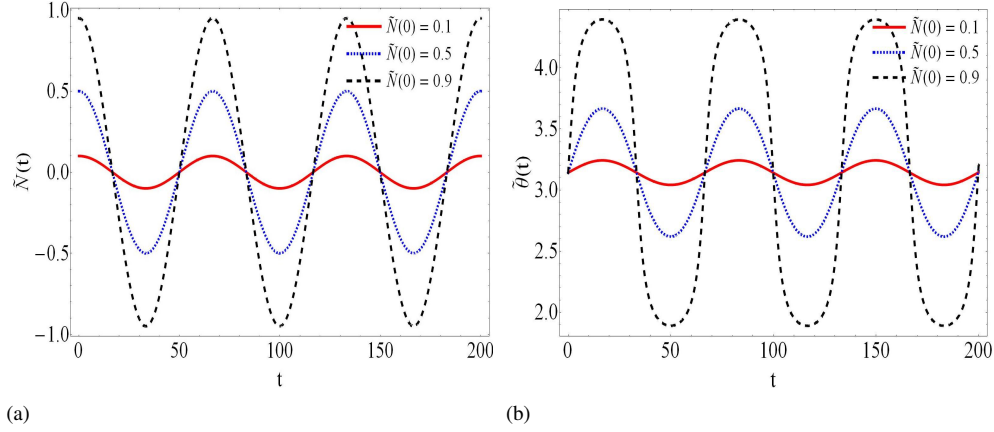

Fig. 3. (a) Fractional population difference $\widetilde{N}(t)$ and (b) the phase difference $\widetilde{\theta}(t)$ per site evaluated as a function of a rescaled time using different initial values of $\widetilde{N}(t)$ and fixed initial value of $\widetilde{\theta}(0) = \pi$. Lattice parameters are $\delta = 0$, $B_{z-bias} = 0\ G$, $\tau = 2\ \mu m$ and $\alpha_s = \alpha_h = 3.5\ \mu m$.

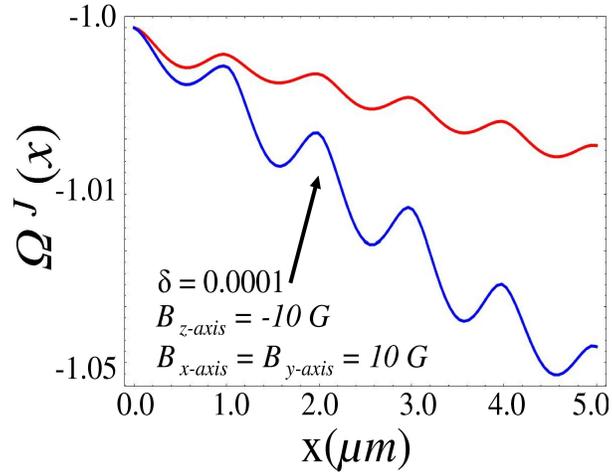

Fig. 4. Numerically evaluated coupling Josephson energy $\Omega^J_{i,i+1}$ with a tilted potential energy U(**x**) ($\delta \neq 0$ and $B_{x-bias} = B_{y-bias} = -B_{z-bias} = 10\ G$) and without a tilted potential energy ($\delta = 0$, $B_{x-bias} = B_{y-bias} = 10\ G$ and $B_{z-bias} = 0\ G$). $\tau = 2\ \mu m$ and $\alpha_s = \alpha_h = 3.5\ \mu m$.

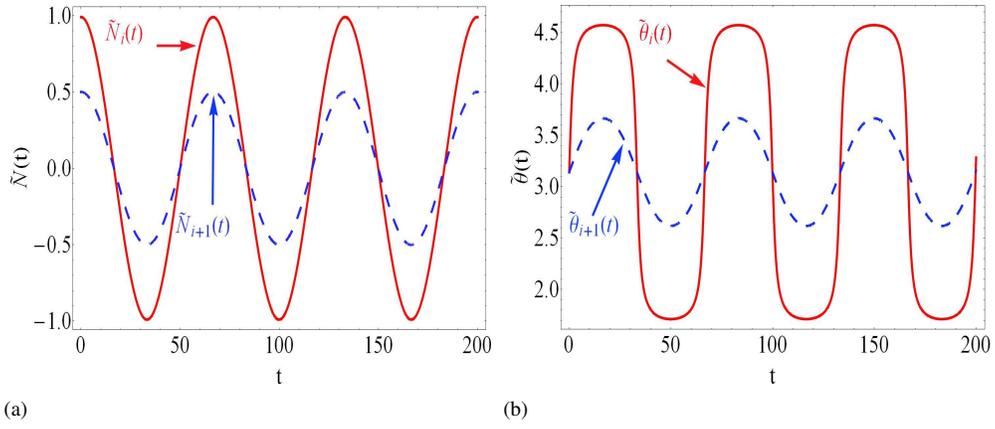

Fig. 5. (a) $\widetilde{N}_i(t)$ and $\widetilde{\theta}_i(t)$ evaluated with time and (b) $\widetilde{N}_{i+1}(t)$ and $\widetilde{\theta}_{i+1}(t)$ with initial values for the ($i$)-site fractional population $\widetilde{N}_i(0) = 0.99$ and for the ($i+1$)-site $\widetilde{N}_{i+1}(0) = 0.1$, while the initial phase differences in both sites are equal, $\widetilde{\theta}_i(0) = \widetilde{\theta}_{i+1}(0) = \pi$. Lattice parameters are $\delta = 0$, $B_{z-bias} = 0\ G$, $\tau = 2\ \mu m$ and $\alpha_s = \alpha_h = 3.5\ \mu m$.